\documentclass[sigplan,screen,nonacm]{acmart}
\AtBeginDocument{%
  }

\usepackage{minted}
\usepackage{booktabs}
\usepackage{multirow}
\usepackage{colortbl}
\usepackage{pifont}
\usepackage{siunitx}
\usepackage{listings}
\usepackage{xcolor}
\usepackage[most]{tcolorbox}

\usepackage{algorithm}
\usepackage{algpseudocode}
\usepackage{amsmath}

\usepackage{amssymb}

\usepackage{fvextra} 

\newcommand{\promptboxfont}{\footnotesize\ttfamily} 

\DefineVerbatimEnvironment{promptverbatim}{Verbatim}{
  formatcom=\promptboxfont,  
  obeytabs=true, tabsize=2,  
  breaklines=true, breakanywhere=true 
}

\newtcolorbox{promptbox}{
  colback=gray!8,
  colframe=blue!60!black,
  arc=3pt,
  boxrule=0.8pt
}

\setcopyright{none}
\copyrightyear{2026}
\acmYear{2026}
\acmDOI{}

\acmConference[KDD '26]{Proceedings of the 32nd ACM SIGKDD Conference on Knowledge Discovery and Data Mining}
{August 9--13, 2026}{Jeju, Republic of Korea}

\acmISBN{}

\begin{document}

\title[Self-Evolved Learning for NL2MQL Generation]{Draft--Refine--Optimize: Self-Evolved Learning for Natural Language to MongoDB Query Generation}

\author{Mingwei Ye}
\authornote{Both authors contributed equally to this research.}
\email{yemingwei@dp.tech}
\affiliation{%
  \institution{DP Technology}
  \state{Beijing}
  \country{China}
}

\author{Jiaxi Zhuang}
\authornotemark[1]
\email{zhuangjiaxi@dp.tech}
\affiliation{%
  \institution{DP Technology}
  \state{Beijing}
  \country{China}
}

\author{Mingjun Xu}
\email{xumj@dp.tech}
\affiliation{%
  \institution{DP Technology}
  \city{Beijing}
  \country{China}
}

\author{Linfeng Zhang}
\affiliation{%
 \institution{DP Technology}
 \city{Beijing}
 \country{China}}

\author{Guolin Ke}
\affiliation{%
  \institution{DP Technology}
  \city{Beijing}
  \country{China}}

\author{Hengxing Cai}
\affiliation{%
  \institution{DP Technology}
  \city{Beijing}
  \country{China}}
\email{caihengxing@dp.tech}

\renewcommand{\shortauthors}{Ye et al.}

\begin{abstract}
    Natural Language to MongoDB Query Language (NL2MQL) is essential for democratizing access to modern document-centric databases. Unlike Text-to-SQL, NL2MQL faces unique challenges from MQL's procedural aggregation pipelines, deeply nested schemas, and ambiguous value grounding. Existing approaches use static prompting or one-shot refinement, which inadequately model these complex contexts and fail to systematically leverage execution feedback for persistent improvement.
    We propose \textbf{EvoMQL}, a self-evolved framework that unifies evidence-grounded context construction with execution-driven learning through iterative \textbf{D}raft-\textbf{R}efine-\textbf{O}ptimize (DRO) cycles. Each cycle uses draft queries to trigger query-aware retrieval, dynamically building compact evidence contexts that resolve schema ambiguities and ground nested paths to concrete values. The model then undergoes online policy optimization with execution-based rewards and curriculum scheduling, with refined models feeding back into subsequent cycles for progressive evolution.
    Overall, EvoMQL achieves state-of-the-art execution accuracy of 76.6\% on the EAI in-distribution benchmark and 83.1\% on the TEND out-of-distribution benchmark, outperforming the strongest open-source baselines by up to 9.5\% and 5.2\%, respectively. With only 3B activated parameters, this closed-loop paradigm enables scalable, continuous improvement of NL2MQL systems in production.
\end{abstract}




\maketitle

\section{Introduction}

Natural Language to Query (Text2Query) have long been a practical pathway to democratize data access, enabling non-experts to retrieve, aggregate, and troubleshoot data without mastering query languages~\cite{shi-etal-2025-gen,ma2025sql,liu2025multitend}. While most prior work has focused on Text-to-SQL for relational databases~\cite{xiyansql_pre,shi-etal-2025-gen,shen2024astschemapruning}, modern production data stacks increasingly rely on semi-structured and document-centric storage to accommodate heterogeneous records such as logs, events, user behaviors, and content metadata. In this setting, MongoDB~\cite{natural-language-to-mongosh-dataset} and similar document databases expose expressive query capabilities through MongoDB Query Language (MQL), especially its aggregation pipeline, which composes multiple stages to filter, transform, group, and join collections in a procedural, stage-wise manner. This design naturally fits nested fields and arrays, but it also raises the barrier for human authors: writing correct MQL often requires careful handling of JSON-path grounding, unwinding and regrouping array structures, and debugging subtle stage interactions~\cite{maamari2024deathschemalink,shen2024astschemapruning}. As a result, a reliable NL2MQL interface is not merely a convenience feature—it directly affects the efficiency of analytics workflows, operational diagnosis, and rapid product iteration in document-database-backed systems~\cite{liu2025multitend}. Importantly, NL2MQL cannot be treated as a straightforward variant of Text-to-SQL: the target language is pipeline-structured rather than declarative, schemas are often implicit or only partially specified~\cite{maamari2024deathschemalink}, and failure modes frequently arise from stage composition and nested-path ambiguity rather than table-column selection alone. These characteristics make NL2MQL a distinct, industrially relevant Text2Query problem that calls for dedicated modeling and optimization, rather than direct reuse of Text-to-SQL recipes~\cite{wang2018robust}.

Compared to Text-to-SQL, NL2MQL poses a qualitatively different set of challenges that go beyond surface-level syntax translation. First, MQL's aggregation pipeline is inherently compositional and non-local: correct generation requires not only grounding mentions to potentially nested JSON paths and values, but also constructing a globally consistent sequence of stages whose interactions are correct (e.g., stage ordering constraints, cardinality changes induced by array unwinding, and side effects introduced by joins and projections). Second, despite strong linguistic priors, general-purpose LLMs are not trained to reliably internalize organization-specific schemas and field semantics, and thus frequently hallucinate non-existent paths/values or omit critical stages when supervision and feedback are limited. Third, while recent Text-to-NoSQL benchmarks have started to formalize executable evaluation, existing approaches still largely rely on one-shot prompting or lightweight test-time fixes~\cite{liu2025multitend, DBLP:journals/corr/abs-2502-11201}, leaving two practical gaps: (i) evidence-grounded refinement that systematically retrieves and aligns schema/value/path evidence to resolve grounding ambiguities, and (ii) learning from execution that turns verifiable outcomes into persistent learning signals rather than transient debugging cues. These observations lead to a central challenge we address in this work: can we design an NL2MQL system that couples retrieval-based refinement with execution-driven online optimization in a closed loop, so that each interaction produces higher-quality queries and, crucially, yields trajectories that improve the model over time?

To address the above challenges, we propose \textbf{EvoMQL}, a novel \textbf{Self-Evolved Model-in-the-Loop (MIL)} framework tailored for NL2MQL, together with a corresponding online reinforcement learning paradigm. Unlike conventional one-shot and static RL pipelines, EvoMQL treats inference enhancement and model training as a continuously iterative co-evolution process, forming a closed-loop “\textbf{Draft query generation -- Retrieval-based Refinement -- Online policy optimization (DRO)}” cycle.

Within each MIL iteration, the model first produces a draft MQL, which serves as an intermediate representation to trigger \emph{query-aware} refinement for context construction. Concretely, we expose the database through two parallel schema formats and then perform schema linking and value grounding conditioned on the draft to retrieve compact, high-relevance evidence about nested paths and values. This evidence improves pipeline-level composition and grounding at inference time, and also yields more diverse context variants that strengthen and stabilize subsequent online policy optimization. During policy optimization, we apply GSPO to effectively optimize the MoE architecture; meanwhile, we design a straightforward reward system tailored for structured queries and introduce a curriculum-based data scheduling strategy to stabilize and accelerate training.

Furthermore, EvoMQL establishes a cross-round self-evolution mechanism: after each training round, the updated model is fed back into the next MIL cycle, allowing the refinement components to become increasingly accurate as the model improves. Combined with online curriculum learning, where sample difficulty is re-estimated by the latest model and the training distribution is dynamically re-balanced, the learning process is always aligned with the model's current capability frontier. After three self-evolution rounds, EvoMQL achieves 9.5\% and 5.2\% higher execution accuracy over the strongest open-source baseline on in-distribution (ID) and out-of-distribution (OOD) settings, respectively. Our contributions are summarized as follows:

\begin{itemize}
    \item We propose \textbf{EvoMQL}, a self-evolved \emph{Model-in-the-Loop} framework for NL2MQL that explicitly targets nested-path grounding and pipeline-level compositionality in MongoDB aggregation, bridging a critical gap in existing NL2MQL research.
    
    \item We develop a \textbf{query-aware retrieval and refinement} suite that leverages draft MQL to construct compact, high-relevance evidence contexts, improving both test-time generation and the quality and diversity of learning trajectories.
    
    \item We introduce an \textbf{iterative online reinforcement learning} paradigm in which, after each policy update, the latest model is reused to rerun MIL data collection and reschedule training via an online curriculum. This closed-loop iteration continually refreshes supervision signals and further unlocks model capability beyond one-shot training.
    
    \item We conduct extensive experiments under both ID and OOD settings, with thorough ablation studies isolating the contribution of each component. Results demonstrate that EvoMQL achieves \textbf{state-of-the-art} performance among open-source models on general NL2MQL benchmarks.
\end{itemize}

\section{Related Work}

\subsection{Text-to-Executable Query}
Text-to-SQL is the most widely studied form of text-to-executable query generation, and recent analyses suggest that performance is often dominated by \emph{grounding} and \emph{verification} rather than pure decoding.
First, \emph{schema grounding} (schema linking) is a fundamental bottleneck: oracle experiments demonstrate that more accurate linking substantially lifts the upper bound of end-to-end accuracy~\cite{lei-etal-2020-examining}, and scalability becomes central in real-world large-schema, multi-database environments~\cite{wang-etal-2025-linkalign}.
Second, \emph{value grounding} requires aligning natural-language mentions to concrete database instances and compatible fields; structure-grounded formulations decompose grounding into column grounding, value grounding, and column--value mapping~\cite{deng-etal-2021-structure}, while recent LLM-era approaches increasingly adopt iterative grounding to bridge questions and schemas (e.g., generate-ground-regenerate)~\cite{shi-etal-2025-gen}.
Third, executable queries enable \emph{execution-aware verification}: execution-guided decoding integrates runtime feedback into generation to filter invalid or inconsistent candidates~\cite{wang2018robust}, and recent LLM reasoning systems further combine verification with structured reasoning for stronger robustness~\cite{he-etal-2025-star}.
These challenges are amplified in NL2MQL (e.g., MongoDB aggregation pipelines) due to semi-structured nested paths, arrays, and stage-wise composition; correspondingly, emerging Text-to-NoSQL benchmarks such as TEND/MultiTEND highlight the need for tailored grounding and stage-level correctness mechanisms beyond directly reusing Text-to-SQL pipelines~\cite{lu2025bridging,liu2025multitend}.
However, existing methods are still largely tailored to relational SQL grounding, and remain insufficient for NL2MQL where grounding targets shift to nested JSON paths and structural failures arise from pipeline composition; while TEND/MultiTEND establish executable benchmarks for Text-to-NoSQL, they leave a gap in unifying evidence retrieval, refinement, and verifiable feedback in one loop. We fill this gap by using draft queries as intermediate structures to trigger schema/value/path evidence retrieval for grounded refinement, better matching the unique grounding and compositional challenges of NL2MQL.

\begin{figure*}[t]
  \centering
  \includegraphics[width=\textwidth]{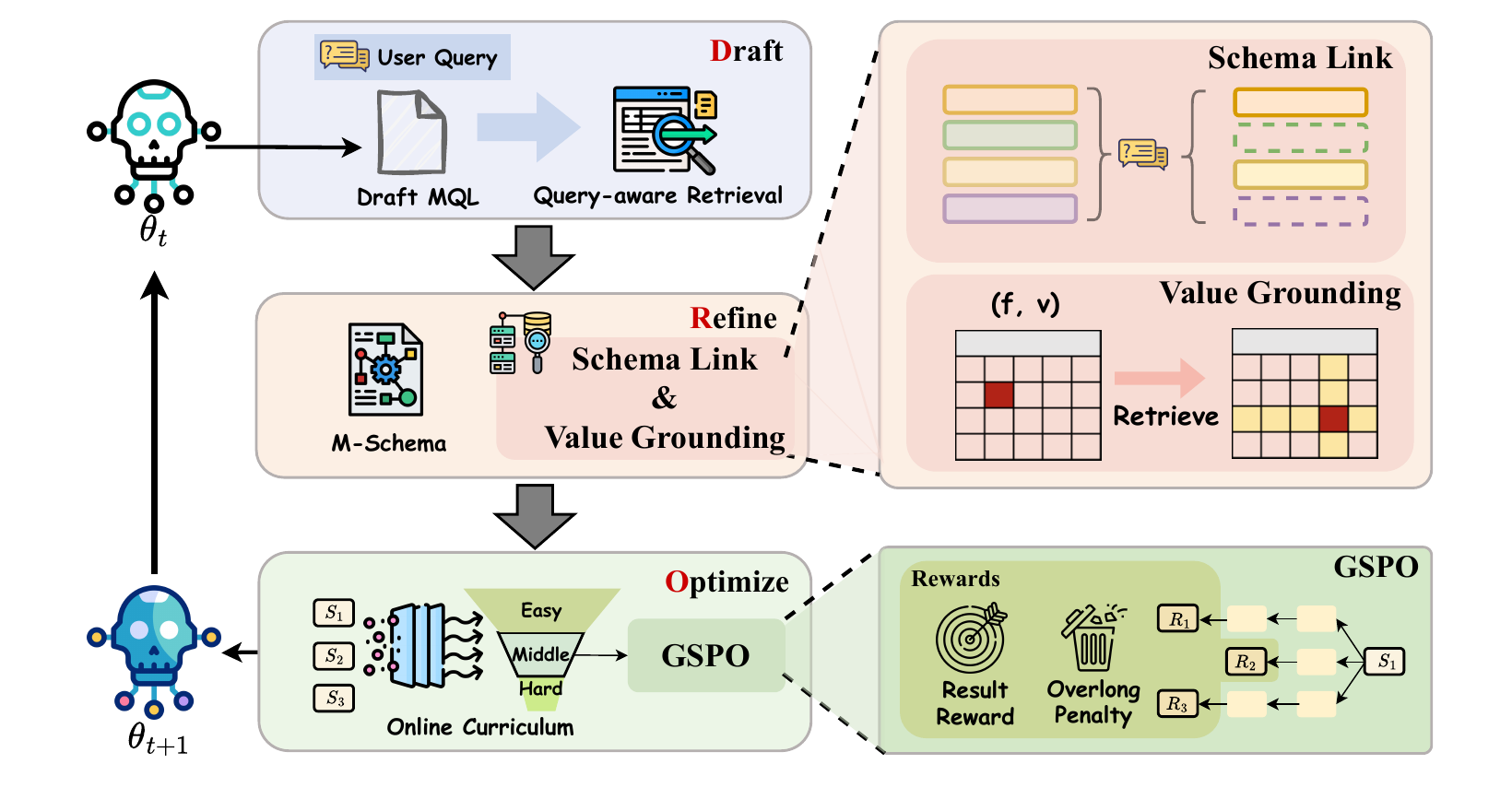}
  \caption{Overall Framework of EvoMQL.}
  \label{fig:evomql-framework}
  \Description{EvoMQL Structure}
\end{figure*}

\subsection{Agentic Refinement and Learning}
With large language models, prompting-based approaches become strong baselines for structured query tasks, but one-shot prompting remains brittle under large schemas and ambiguous values.
This motivates \emph{iterative refinement} and \emph{agentic} pipelines that repeatedly draft a query, consult external evidence (e.g., schema/value lookup or database probing), and revise the query conditioned on feedback.
ReAct provides a generic foundation by interleaving explicit reasoning with tool use in a \textit{Think--Act--Observe} loop, turning LLMs into interactive problem solvers~\cite{yao2022react}.
In text-to-query settings, such agentic prompting is commonly instantiated as execution-driven refinement (draft $\rightarrow$ execute $\rightarrow$ fix), where execution outcomes serve as explicit signals for debugging and correction.
Beyond test-time refinement, a closely related line uses feedback as a \emph{learning signal}, via reinforcement learning (RL) or self-improvement loops.
For text-to-query, RL must handle sparse and easily hackable rewards; thus, representative methods exploit execution-aware optimization and verification-driven supervision to stabilize training and improve semantic correctness~\cite{dai2025reex,he-etal-2025-star}.
Recent group-based policy optimization methods offer practical building blocks for such online learning.
GRPO (introduced in DeepSeekMath) optimizes policies using group-relative advantages without an explicit critic, improving memory efficiency~\cite{shao2024deepseekmath}.
GSPO further improves stability by using sequence-level policy optimization with clipping, and has been adopted for scalable RL post-training recipes~\cite{zheng2025gspo}.
In contrast to prior work that focuses mainly on test-time refinement, and RL/self-training approaches mostly studied on Text-to-SQL without tightly coupling evidence retrieval, we unify Draft, retrieval-based Refine, and online Optimize into a closed-loop pipeline (Draft $\rightarrow$ Refine $\rightarrow$ Optimize), so that interaction trajectories (retrieved evidence + execution outcomes) become direct learning signals for continual improvement in NL2MQL.

\section{Method}

\subsection{Overview}
EvoMQL transforms the static NL2MQL training pipeline into a dynamic, Self-Evolved Model-in-the-Loop (MIL) architecture. As illustrated in Figure~\ref{fig:evomql-framework}, the framework adopts an iterative closed loop to bridge natural-language intent and complex MQL syntax. Starting from the current policy $\theta_t$, the model first generates a draft MQL. Based on this draft, three refinement strategies are applied to construct a more information-dense context. Leveraging the expanded reasoning trajectories, we estimate the difficulty distribution, organize curriculum-aware training data, and optimize the policy using GSPO. The resulting model $\theta_{t+1}$ is then fed into the next iteration, enabling progressive self-evolution. Detailed components are described in the following sections.

\subsection{Draft}

\subsubsection{Schema Design}

Schema representation plays a critical role in NL2Query tasks, as it constitutes the primary channel through which large language models perceive the structure and semantics of a database. In the Atlas Benchmark, schemas are provided in a TypeScript format that describes the key elements of a database, including collection-level descriptions, field names, field types, field descriptions, and a small set of example values. For MongoDB in particular, schema representations must additionally account for nested fields (i.e., hierarchical children structures) as well as index information, which is essential for understanding query efficiency and feasible access patterns.

In addition to the original TypeScript schema, we build upon \textbf{M-Schema}~\cite{xiyansql_pre}, a structured and information-dense schema format originally proposed for NL2SQL. We further adapt M-Schema to MongoDB's semi-structured setting by explicitly preserving nested-field hierarchy and incorporating MongoDB-specific metadata when available (e.g., index information) to better reflect feasible access patterns. In our framework, the TypeScript schema and M-Schema are treated as two parallel schema formats used for context construction. Figure~2 illustrates both.

\subsubsection{Draft Generation}

Given a schema $S$ and a natural language query $Q$, the current policy $\theta_t$ first produces an executable draft MQL $M_d$. We prompt the model to reason explicitly, while requiring the final MQL to be returned as a stage-delimited pipeline inside a JavaScript Markdown code block, enabling reliable parsing of operators, fields, and values.

Using $M_d$ as a query-shaped anchor, our refinement modules can precisely diagnose missing evidence and ambiguities, retrieve compact schema/value support, and construct an information-dense context that reduces hallucinated paths and improves pipeline-level compositionality for downstream generation and optimization.

\subsection{Refinement}

\subsubsection{Schema Linking}

Introducing excessive or irrelevant schema elements into the prompt can significantly degrade generation quality~\cite{maamari2024deathschemalink,shen2024astschemapruning}. First, it increases the likelihood that the model will reference irrelevant collections or fields during query generation. Second, providing the full database schema often results in overly long prompts, which may exceed the model's context window or incur unnecessary inference costs~\cite{liu2024lostmiddle,du2025contextlength}.

To address these issues, we perform schema linking in two steps: (i) a high-precision rule-based extraction from the draft MQL $M_d$, and (ii) a recall-oriented semantic expansion over the full schema $S$.

\textbf{Step 1: rule-based field extraction.}
We first parse $M_d$ and extract all field paths that are explicitly referenced by any stage/operator (e.g., $\$match$/$\$group$/$\$project$) to obtain an initial field set

\begin{equation}
  \mathcal{F}_0 = \mathrm{ExtractFields}(M_d).
\end{equation}

In practice, $\mathrm{ExtractFields}(\cdot)$ is implemented with lightweight rules over the JSON-like AST of the draft query. This step is intentionally conservative: it prioritizes precision to avoid introducing irrelevant schema noise.

\textbf{Step 2: semantic expansion via embeddings.}
Because $M_d$ can be incomplete or partially incorrect, $\mathcal{F}_0$ may miss fields that are semantically implied by the question but not surfaced in the draft. We therefore expand each extracted field $f \in \mathcal{F}_0$ by retrieving semantically related fields from the schema using an embedding model $\phi(\cdot)$ and a vector index built over all schema fields (including nested paths and their descriptions when available). Concretely, we compute
\begin{equation}
  \mathcal{R}_k(f)=\operatorname{TopK}_{g\in\mathcal{U}(S)}\; \mathrm{sim}(\phi(f),\phi(g)),
\end{equation}
where $\mathcal{U}(S)$ denotes the universe of candidate fields in $S$, and $\mathrm{sim}(\cdot,\cdot)$ is cosine similarity. The expanded field set is then
\begin{equation}
  \mathcal{F}=\mathcal{F}_0 \cup \bigcup_{f\in\mathcal{F}_0} \mathcal{R}_k(f).
\end{equation}
Finally, we construct the pruned schema $S'$ by projecting $S$ onto $\mathcal{F}$. This procedure keeps the prompt compact while recovering semantically relevant fields that the draft may omit, striking a balance between noise reduction and coverage.

\subsubsection{Value Grounding}

Structured queries frequently involve matching concrete values, while natural language descriptions of such values may be ambiguous, imprecise, or even incorrect~\cite{huang2023dataambiguity,talaei2024chess,deng-etal-2021-structure,brunner-stockinger-2021-valuenet}. Moreover, when multiple semantically similar fields exist, the model's search space can become misaligned, leading to inaccurate value grounding in a single-shot generation~\cite{bhaskar2023ambiqt,talaei2024chess,deng-etal-2021-structure,brunner-stockinger-2021-valuenet}.

\textbf{Value grounding} aims to align mentions in $Q$ with concrete database instances and their most compatible fields. Similar to schema linking, we start from the draft query $M_d$ and perform a high-precision extraction followed by a recall-oriented expansion.

\textbf{Step 1: rule-based predicate extraction.}
We parse $M_d$ and extract all value-matching predicates (e.g., equality, range constraints, membership tests) to obtain an initial set of field--value pairs
\begin{equation}
  \mathcal{V}_0 = \mathrm{ExtractValues}(M_d)=\{(f, v)\}.
\end{equation}

\textbf{Step 2: probe-based verification.}
Because the field in $\mathcal{V}_0$ may be incorrect or overly specific, we reuse the semantic expansion operator $\mathcal{R}_k(\cdot)$ defined in Eq.~(3) to construct a candidate field set $\mathcal{F}_c$.
For each candidate $g\in\mathcal{F}_c$, we issue a lightweight MQL query to test whether the value $v$ occurs under field $g$. If the probe returns at least one match, we keep $(g,v)$ and record the returned evidence. Formally, we collect
\begin{equation}
  \mathcal{E}(f,v)=\{e_g\mid g\in\mathcal{F}_c,\; \mathrm{Match}(g,v)=1\},
\end{equation}
where $e_g$ denotes the probe evidence for $(g,v)$. We provide $\mathcal{E}(f,v)$ as auxiliary evidence, enabling the LLM to better resolve the final field--value alignment.

\subsubsection{Summary}

Overall, our refinement pipeline systematically constructs \emph{more accurate and actionable evidence contexts} for NL2MQL generation. By retrieving schema- and value-level evidence conditioned on draft queries, the resulting context helps the model better internalize the underlying MongoDB schema, precisely ground nested paths, and identify the fields and concrete values that are truly relevant to the user question. In addition, independently enabling the three refinement modules yields up to $2^3 = 8$ context variants per instance, increasing training-time context diversity and improving prompt robustness for subsequent policy optimization.

\subsection{Reinforcement Learning}

\subsubsection{Reward Design}
\label{sec:reward-design}

Since structured query generation naturally fits the RLVR (Reinforcement Learning with Verifiable Rewards) paradigm, and RLVR has been empirically validated across multiple NL2SQL benchmarks, we adopt this optimization framework for NL2MQL as well~\cite{ma2025sql}. Crucially, MQL queries admit deterministic execution and result-level verification, which allows us to construct dense, reliable, and non-heuristic rewards without relying on costly human annotation. Accordingly, we design two reward components: \textbf{Result Reward}, and \textbf{Overlong Penalty}.

\textbf{Result Reward.}
The Result Reward evaluates the overall correctness of the generated MQL, including both syntactic validity and execution-level result correctness. Following~\cite{yao2025arctic}, we assign a strong negative reward when the model fails to produce a valid MQL query or when the query cannot be executed. If the query is syntactically valid but yields incorrect results, we provide a small positive reward to encourage partial progress. When the query returns correct results, we further distinguish whether the output is overly permissive. Due to the fuzzy-matching protocol used in evaluation (see \S\ref{sec:metrics}), returning too many fields can trivially satisfy the metric, leading to reward hacking. Therefore, such cases are only weakly rewarded.
Formally, the Result Reward is defined as
\begin{equation}
R_{\mathrm{res}}(\hat{y})=
\begin{cases}
-1.0, & \text{if } \hat{y} \text{ syntax error},\\
0.1, & \text{elif } \hat{y} \text{ produces incorrect results},\\
0.25, & \text{elif } \mathrm{Correct}(\hat{y}) \wedge \mathrm{OverField}(\hat{y}),\\
1.0, & \text{else}.
\end{cases}
\end{equation}
Here, $\hat{y}$ denotes the generated MQL query, and $y$ denotes the reference (ground-truth) query. $\mathcal{F}(\hat{y})$ denotes the set of returned fields. We define $\mathrm{OverField}(\hat{y})=\mathbb{I}[|\mathcal{F}(\hat{y})|>|\mathcal{F}(y)|+\tau]$, where $\tau$ is a threshold controlling excessive field inclusion.

\textbf{Overlong Penalty.}
To improve both training stability and inference efficiency, and motivated by recent findings that excessively long reasoning chains can harm smaller models~\cite{luo2025valley,li2025learnabilitygap,wu2025moreless}, we introduce an Overlong Penalty based on DAPO~\cite{yu2025dapo}. This term penalizes responses that exceed a soft length budget while keeping a small buffer for necessary reasoning.
The penalty is defined as
\begin{equation}
R_{\text{over}}(y)=
\begin{cases}
0, & |y|\le L_{\max}-L_{\mathrm{cache}},\\[6pt]
-\lambda\cdot\dfrac{|y|-(L_{\max}-L_{\mathrm{cache}})}{L_{\mathrm{cache}}},
& L_{\max}-L_{\mathrm{cache}}<|y|\le L_{\max},\\[10pt]
-\lambda, & |y|>L_{\max}.
\end{cases}
\end{equation}
This formulation enforces a smooth penalty near the maximum length while strongly discouraging pathological over-generation.

Finally, we define the total reward as the sum of the two components:
\begin{equation}
R(\hat{y}) = R_{\mathrm{res}}(\hat{y}) + R_{\text{over}}(\hat{y}).
\end{equation}

\subsubsection{Policy Optimization}

We employ GSPO for policy optimization which maximizes the expected reward across rollout trajectories:
\begin{equation}
\begin{aligned}
\mathcal{J}_{\text{GSPO}}(\theta)
&= \mathbb{E}_{x \sim \mathcal{D},\; \{y_i\}_{i=1}^G} \Bigg[ \\
&\qquad \frac{1}{G} \sum_{i=1}^{G}
\min \Big(
    s_i(\theta)\,\hat{A}_i,\;
    \text{clip}(s_i(\theta))\,\hat{A}_i
\Big)
\Bigg].
\end{aligned}
\end{equation}
Here, $s_i(\theta)$ denotes the sequence-level importance ratio, defined as the geometric mean of token-wise likelihood ratios:
\begin{equation}
\begin{aligned}
s_i(\theta)
&= \left( \frac{\pi_{\theta}(y_i \mid x)}{\pi_{\theta_{\text{old}}}(y_i \mid x)} \right)^{\frac{1}{|y_i|}}\\
&= \exp \left( \frac{1}{|y_i|} \sum_{t=1}^{|y_i|} \log \frac{\pi_{\theta}(y_{i,t} \mid x, y_{i,<t})} {\pi_{\theta_{\text{old}}}(y_{i,t} \mid x, y_{i,<t})} \right).
\end{aligned}
\end{equation}
The sequence-level ratio and group-based advantage formulation in GSPO provide a more stable optimization signal for MoE policies with heterogeneous expert activations.

\subsection{Online Curriculum Update}
\label{sec:online-update}

EvoMQL performs cross-iteration self-evolution by repeatedly executing the DRO loop. Starting from the current policy $\pi_{\theta_t}$, each iteration (i) drafts an intermediate MQL query, (ii) builds a refined, evidence-grounded prompt via retrieval-based refinements, and (iii) optimizes the policy with online reinforcement learning under our reward design, producing an updated policy $\pi_{\theta_{t+1}}$. The updated policy then serves as the initialization for the next iteration, enabling progressive improvement.

A key component for stabilizing this online learning process is \emph{difficulty-aware data scheduling}. The motivation is to keep training focused on examples that are neither trivial nor hopelessly hard---i.e., within the model's ``proximal zone of development''~\cite{vygotsky1978mind}---which is closely related to curriculum learning principles~\cite{zhang-etal-2025-learning-like}.

Concretely, after constructing refinement-augmented contexts for the current iteration and right before policy optimization, we re-evaluate the difficulty of each training instance under the latest policy. For each instance $x_i$, we perform $N$ independent rollouts with $\pi_{\theta_t}$ and compute a binary correctness indicator $\mathbb{I}_i^{(j)}$ based on fuzzy execution matching. We define the instance difficulty as
\begin{equation}
  d_i = N - \sum_{j=1}^{N} \mathbb{I}_i^{(j)}.
\end{equation}
Intuitively, $d_i=0$ means the instance is consistently solved, while $d_i=N$ indicates consistent failure.

Using these difficulty estimates, we apply a simple band-pass filter and keep only instances with moderate difficulty:
\begin{equation}
  d_{\min} < d_i < d_{\max}.
\end{equation}
This selection mitigates gradient saturation from easy examples and reduces noise from outliers that are currently unsolvable.

Finally, we implement the cross-iteration curriculum with a \emph{candidate pool} and a \emph{progressive retention schedule}. At iteration $t$, we first apply the policy-dependent band-pass filter to obtain a moderate set $\mathcal{D}^{\mathrm{mid}}_t=\{x_i\mid d_{\min}<d_i<d_{\max}\}$, then sample a fraction $\rho_t$ of $\mathcal{D}^{\mathrm{mid}}_t$ for policy optimization. The selected instances are removed from the candidate pool so that later iterations focus on newly challenging examples under the improved policy, and difficulty is re-estimated by the updated model before the next round.

\begin{table*}
\caption{Main results on ID and OOD benchmarks. The best results are highlighted in \textbf{bold}, and the second-best results are \underline{underlined}. Our method is highlighted with a \colorbox{gray!15}{\strut\textbf{gray}} background.}
\label{tab:main_results}
\centering
\resizebox{\textwidth}{!}{%
\begin{tabular}{lccccc|ccccc}
\toprule
\multicolumn{1}{c}{\multirow{2.5}{*}{\textbf{Model}}} 
& \multicolumn{5}{c|}{\textbf{EAI Benchmark (ID)}} 
& \multicolumn{5}{c}{\textbf{TEND Benchmark (OOD)}} \\
\cmidrule(lr){2-6} \cmidrule(lr){7-11}
& SE & NEO & RO & \textbf{COF} & \textbf{OPS} 
& SE & NEO & RO & \textbf{COF} & \textbf{OPS} \\

\midrule

GPT-5          & \underline{0.952} & 0.889 & \textbf{0.872} & 0.594 & 0.723 
               & \textbf{1.000} & \textbf{0.987} & \textbf{0.987} & 0.805 & 0.880 \\
GPT-4o         & 0.938 & 0.907 & \underline{0.853} & 0.700 & 0.784
               & \underline{0.987} & 0.961 & 0.935 & 0.831 & 0.886 \\
Gemini-3-flash              & 0.935 & 0.883 & 0.792 & 0.734 & 0.795
               & 0.935 & 0.935 & 0.935 & \textbf{0.909} & \underline{0.919} \\
Gemini-3-pro                & 0.949 & 0.894 & 0.834 & 0.721 & 0.795
               & 0.974 & \underline{0.974} & \underline{0.974} & \underline{0.896} & \textbf{0.927} \\

\midrule

MiniMax-M2                    & 0.949 & 0.855 & 0.726 & 0.666 & 0.748 
               & 0.961 & 0.961 & 0.844 & 0.779 & 0.840 \\
gpt-oss-20b                   & 0.660 & 0.634 & 0.598 & 0.457 & 0.529
               & 0.909 & 0.909 & 0.896 & 0.766 & 0.822 \\
Llama-3.1-8B-Instruct         & 0.798 & 0.616 & 0.519 & 0.418 & 0.524 
               & 0.779 & 0.649 & 0.558 & 0.338 & 0.479 \\
Qwen3-Coder                   & 0.926 & 0.890 & 0.780 & 0.671 & 0.755 
               & 0.909 & 0.909 & 0.844 & 0.688 & 0.770 \\
Qwen3-30B-A3B-Thinking        & 0.913 & 0.879 & 0.729 & 0.671 & 0.746 
               & \underline{0.987} & 0.909 & 0.766 & 0.740 & 0.809 \\

\rowcolor{gray!15}
\textbf{EvoMQL (Draft)} & \underline{0.952}	& \underline{0.921} & 0.749 & \underline{0.742} & \underline{0.803}
                & 0.974 & 0.922 & 0.844 & 0.792 & 0.847 \\
\rowcolor{gray!15}
\textbf{EvoMQL} & \textbf{0.963} & \textbf{0.930} & 0.753 & \textbf{0.766} & \textbf{0.821}
                & \underline{0.987} & 0.935 & 0.792 & 0.831 & 0.869 \\

\bottomrule
\end{tabular}}
\end{table*}
In our experiments we run a fixed number of DRO iterations (three rounds). Concrete hyperparameters ($N$, $d_{\min}$/$d_{\max}$, and $\{\rho_t\}$) are given in \S\ref{sec:curriculum-setting}.

\section{Experiments}

\subsection{Experimental Setup}

\textbf{Datasets.} 
We build our training data by synthesizing a large set of NL2MQL instances from the seven databases in the official MongoDB-EAI NL2Mongosh benchmark (EAI)~\cite{natural-language-to-mongosh-dataset}. The synthesis procedure and implementation details are described in the appendix~\ref{app:data_synthesis}.
We then train the RL policy on the synthesized training set and evaluate on two benchmarks: (i) the \textbf{EAI} benchmark as our in-distribution (ID) test set, and (ii) the MQL-related subset of the \textbf{TEND benchmark}—a more demanding NL2NoSQL dataset—as our out-of-distribution (OOD) test set~\cite{lu2025bridging}. Since TEND covers multiple NoSQL query languages, we specifically focus on its MQL component, which features significantly higher structural complexity with schemas involving up to five levels of nesting. Moreover, it presents a higher linguistic challenge, as the natural-language questions predominantly target complex aggregate operations, requiring sophisticated multi-stage reasoning.

\textbf{Metrics.} 
\label{sec:metrics}
We follow the official evaluation protocol and report four execution-based metrics (with shorthand notations used throughout the paper):
(1) \textbf{Successful Execution (SE)}: the generated query runs without error.
(2) \textbf{Correct Output Fuzzy(COF)}: the query output matches the reference output under the benchmark's fuzzy matching rule, which tolerates minor differences such as extra fields or slightly relaxed \emph{numeric} constraint ranges.
(3) \textbf{Non-Empty Output (NEO)}: the query returns a non-empty result (e.g., excluding [], \{\}).
(4) \textbf{Reasonable Output (RO)}: the returned result is non-empty and contains no null values or empty strings.

To provide a single summary number, we further report an \textbf{Overall Performance Score (OPS)} as a weighted combination of the above metrics:

\begin{equation}
  \text{OPS} = 0.6\cdot \text{COF} + 0.2\cdot \text{SE} + 0.1\cdot \text{NEO}+ 0.1\cdot \text{RO}.
\end{equation}

\textbf{Baselines.} We compare EvoMQL against strong proprietary LLMs, including GPT-5 and Gemini 3, as well as state-of-the-art open-source models such as MiniMax-M2 and Qwen3~\cite{yang2025qwen3}, under the same benchmark setting and evaluation metrics.

\textbf{Implementation Settings.}
\label{sec:curriculum-setting}
EvoMQL is built on Qwen3-30B-A3B-Thinking. For RL training, we use a learning rate of $1\times 10^{-6}$, 8 actor rollouts, and a maximum response length of 4096 tokens. For inference, we set the temperature to 0.7 and top\_p to 0.95.

For reward design (\S\ref{sec:reward-design}), we set $\tau=3$ and $\lambda=0.2$. For online curriculum updates (\S\ref{sec:online-update}), we estimate difficulty with $N=9$ rollouts and use a band-pass filter with $d_{\min}=1$ and $d_{\max}=8$. We run three DRO iterations with progressive retention ratios $\rho=\{0.5, 0.5, 1.0\}$ over the filtered candidate pool.

\subsection{Main Results}

Table~\ref{tab:main_results} compares \textbf{EvoMQL} with representative proprietary LLMs and strong open-source baselines on the \textbf{EAI (ID)} and \textbf{TEND (OOD)} benchmarks. We focus on \textbf{COF} and \textbf{OPS}, as they best capture end-to-end NL2MQL usability.

\textbf{In-distribution performance on EAI.}
EvoMQL achieves the strongest overall performance across all methods, including proprietary commercial models, with \textbf{COF} of \textbf{0.766} and \textbf{OPS} of \textbf{0.821}. It outperforms the strongest non-ours baseline, improving COF by 3.2\% and OPS by 2.6\%. These gains demonstrate that our self-evolved online RL can translate limited model capacity into pipeline-level execution improvements. Notably, we obtain this advantage with only \textbf{$\sim$3B activated parameters}, highlighting the efficiency of the proposed training framework.

\textbf{Out-of-distribution performance on TEND.}
OOD generalization is more challenging, and closed-source frontier models remain highly competitive. Nevertheless, EvoMQL delivers the strongest results among open-source baselines, with \textbf{COF} and \textbf{OPS} reaching \textbf{0.831} (+5.2\%) and \textbf{0.869} (+2.9\%), respectively. EvoMQL also matches GPT series on COF with \textbf{$\sim$3B activated parameters}, demonstrating highly competitive performance with strong robustness--efficiency trade-offs.

\textbf{Effect of RL under identical context.}
To isolate the contribution of our online RL from context augmentation, we also report \textbf{EvoMQL (Draft)}, which uses the same query context without refinement. Even without refinement, RL training still delivers clear improvements and strong absolute performance: on ID it remains ahead of the strongest baseline, and on OOD it continues to outperform the best open-source baseline. These results confirm that the improvements are not merely due to richer prompts, but stem from \emph{training-time policy optimization} that directly targets execution correctness.

\subsection{Ablation Study}

\textbf{Impact of Refinement on Inference.}
As shown in Table~\ref{tab:abl_refine_infer}, all three refinement components contribute to stronger test-time performance. Compared to EvoMQL (Draft), \textit{schema linking} and \textit{value grounding} provide substantial OOD improvements, while \textit{mschema} mainly benefits the ID setting but is less robust under distribution shift. Overall, these results validate that refinement is not a cosmetic prompt augmentation, but a critical mechanism for grounding paths and values and improving pipeline composition during inference.

Beyond improving test-time generation, we next ask whether these same refinement components also provide 
\emph{progressive} benefits during self-evolution by shaping the learning signal and the distribution of training contexts.

\begin{table}[t]
\centering
\begin{tabular}{l
                S[table-format=1.3] S[table-format=1.3]
                S[table-format=1.3] S[table-format=1.3]}
\toprule
\multirow{2}{*}{Methods} 
& \multicolumn{2}{c}{EAI} & \multicolumn{2}{c}{TEND} \\
\cmidrule(lr){2-3} \cmidrule(lr){4-5}
& {COF} & {OPS} & {COF} & {OPS} \\
\midrule
EvoMQL (Draft)   & 0.742 & 0.803 & 0.766 & 0.826 \\
w/ mschema       & 0.755 & 0.813 & 0.727 & 0.801 \\
w/ schema link   & 0.753 & 0.809 & 0.805 & 0.847 \\
w/ value ground  & 0.752 & 0.809 & 0.792 & 0.847 \\
\textbf{EvoMQL}  & \bfseries 0.766 & \bfseries 0.821 & \bfseries 0.831 & \bfseries 0.869 \\
\bottomrule
\end{tabular}
\caption{Ablation results of refinement components during inference on the ID and OOD benchmarks.}
\label{tab:abl_refine_infer}
\end{table}

\textbf{Impact of Refinement on Training.}
To ensure a fair comparison, we strictly control the training budget by matching the number of optimization steps in each iteration across all methods. In addition, to exclude any test-time advantages, all models are evaluated under an identical inference setting, where no refinement augmentation (i.e., Draft MQL) is applied.

As shown in Table~\ref{tab:abl_refinement_on_train}, removing the refinement modules consistently degrades performance, despite identical training compute and identical inference inputs. This observation suggests that refinement is not merely a test-time heuristic, but plays a substantive role during training by shaping more effective learning signals. Furthermore, enabling the three refinement modules independently increases the diversity of training-time contexts, which in turn improves prompt robustness and leads to more stable and effective policy optimization.

\begin{table}[t]
\centering
\begin{tabular}{ccc cc}
\toprule
mschema & schema link & value ground & \textbf{COF} & \textbf{OPS} \\
\midrule
$\checkmark$ & $\checkmark$ & $\checkmark$ & \textbf{0.792} & \textbf{0.847} \\
\midrule
$\times$ & $\times$ & $\times$ & 0.675 & 0.761 \\
$\times$ & $\checkmark$ & $\checkmark$ & 0.753 & 0.827 \\
$\checkmark$ & $\times$ & $\checkmark$ & 0.740 & 0.808 \\
$\checkmark$ & $\checkmark$ & $\times$ & 0.714 & 0.786 \\
\bottomrule
\end{tabular}
\caption{Ablation study of refinement components during training on the OOD benchmark. $\checkmark$/$\times$ indicates whether a module is enabled during \textbf{training}. All results are evaluated under the same training budget and inference input.}
\label{tab:abl_refinement_on_train}
\end{table}

\begin{figure}[t]
  \centering
  \includegraphics[width=\columnwidth]{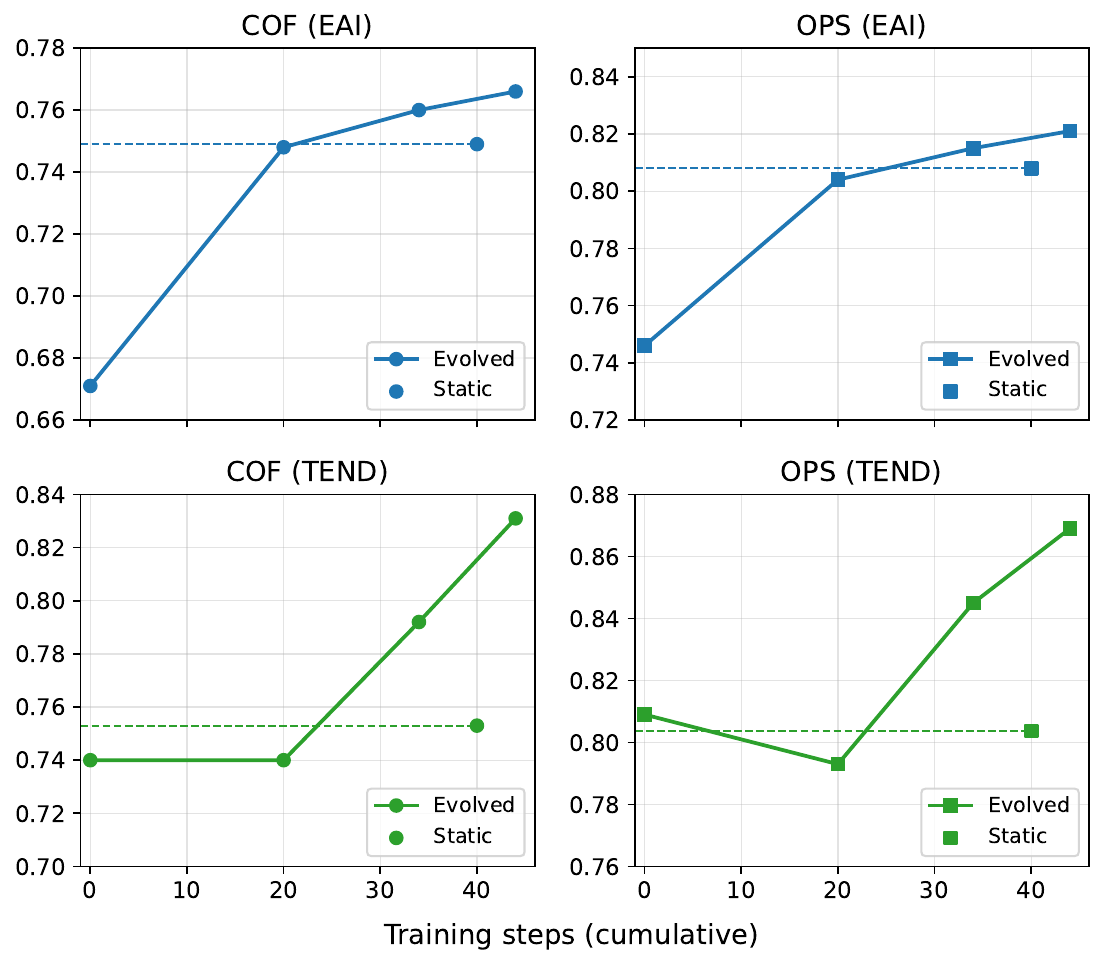}
  \caption{Performance of online curriculum learning over cumulative training steps. We report COF (left) and OPS (right) on EAI (top) and TEND (bottom). Solid lines denote the three iterations of the evolved model, while dashed lines indicate the one-epoch static baseline.}
  \label{fig:evo-analysis}
  \Description{Evo-analy}
\end{figure}
\textbf{Effect of Cross-Iteration Self-Evolution.}
Figure~\ref{fig:evo-analysis} evaluates the benefit of cross-iteration self-evolution by comparing \emph{static} training (a single pass over a fixed training pool) with \emph{evolved} training (three consecutive DRO iterations). The static baseline shows a clear saturation effect: increasing training from 20 steps, which correspond to the first self-evolution iteration, to a full epoch of 40 steps yields only marginal gains on both benchmarks. Because both settings are trained on the same data distribution, this result suggests that the additional updates are largely unproductive due to an unfavorable difficulty distribution.

In contrast, self-evolution consistently produces monotonic gains as iterations proceed, and achieves substantially better final performance with only \emph{negligible} additional optimization steps ($+5\%$ steps) beyond one full data pass. This indicates that the improvement is not merely a consequence of longer training, but rather comes from \emph{refreshing the learning signal} across iterations: the updated policy generates stronger draft queries, which in turn enables more accurate retrieval-based refinements and evidence-grounded prompts, while difficulty-aware scheduling keeps optimization focused on examples within the model's proximal zone of development. Together, these effects stabilize online RL and sustain effective learning beyond the plateau reached by static training.

\section{Conclusion}

In this paper, we present EvoMQL, a self-evolved Model-in-the-Loop framework for NL2MQL that integrates draft query generation, retrieval-based refinement, and online policy optimization into a unified closed-loop. By leveraging retrieval-based context refinement together with iterative online reinforcement learning, EvoMQL effectively addresses the challenges of nested-path grounding and pipeline-level compositionality in MongoDB aggregation queries.

Extensive experiments on both the official EAI MongoDB benchmark and TEND benchmark demonstrate that EvoMQL achieves state-of-the-art execution accuracy while maintaining competitive inference efficiency. These results suggest that tightly coupling query-aware context refinement with online policy evolution is a promising direction for advancing natural-language interfaces to complex document databases.

\section*{Artifact Availability}

To facilitate reproducibility, we release the implementation and experiment scripts used in this paper at \url{https://anonymous.4open.science/r/EvoMQL-F0FB/}.


\newpage
\bibliographystyle{ACM-Reference-Format}

\appendix

\newpage
\section{Data Synthesis Pipeline}
\label{app:data_synthesis}

To address the scarcity of high-quality, executable NL2MQL annotated data, we designed a closed-loop data synthesis pipeline incorporating \textit{Extraction}, \textit{Generation}, and \textit{Verification}. This pipeline leverages real-world MongoDB databases to ensure authentic data distribution and introduces an Execution-Guided Rejection Sampling mechanism to guarantee query correctness.

\subsection{Schema-Aware Context Construction}
Unlike approaches relying solely on schema definitions, we employ a data-grounded context augmentation strategy to minimize hallucinations.

\begin{itemize}
    \item \textbf{Schema Extraction:} We extract the complete structure of collections, including nested fields, type constraints, and index information, to construct a deterministic schema graph.
    \item \textbf{Representative Value Sampling:} To ensure that generated query conditions (e.g., predicates in \texttt{\$match}) are logically valid and executable, we apply a Mixed Sampling Strategy for each collection: 50\% random sampling, 30\% stratified sampling based on categorical fields, and 20\% rare value sampling for numerical fields. This provides the Large Language Model (LLM) with concrete "anchor" values derived from the actual data distribution.
\end{itemize}

\subsection{Complexity-Controlled Generation with CoT}
We utilize GPT-4o as the backbone generator, employing strict prompt engineering to control the distribution of the synthesized data.

\begin{itemize}
    \item \textbf{Chain-of-Thought Reasoning:} We enforce the model to output a reasoning trajectory (enclosed in \texttt{<think>} tags) before generating the final JSON-formatted (NL, MQL) pair. This mechanism not only enhances the logical correctness of the MQL but also preserves reasoning traces for potential Supervised Fine-Tuning (SFT).
    \item \textbf{Dynamic Complexity Distribution:} To simulate real-world analytical scenarios, we define three complexity levels: \textit{Simple} (single-field lookups, 30\%), \textit{Moderate} (multi-condition logical operators, 40\%), and \textit{Complex} (aggregation pipelines with multi-stage processing, 30\%). The system dynamically adjusts the generation constraints based on these target ratios.
\end{itemize}

\subsection{Execution-Guided Rejection Sampling}
The core of our quality assurance lies in validating the generated queries against the actual database. For each natural language intent $x$, we generate a set of $N$ candidate MQL queries $\mathcal{C} = \{y_1, \dots, y_N\}$ (with $N=5 \sim 8$ in our experiments).

Each candidate $y_i$ is executed on a live MongoDB instance. We calculate a quality score $S(y_i)$ based on the execution feedback:

{\small
\begin{equation}
S(y_i) = w_1 \cdot \mathbb{I}_{syn}(y_i) + w_2 \cdot \mathbb{I}_{exec}(y_i) + w_3 \cdot \mathbb{I}_{valid}(R_i) + w_4 \cdot \mathbb{I}_{consist}(x, y_i)
\end{equation}
}

where $\mathbb{I}_{syn}$ denotes syntactic validity, $\mathbb{I}_{exec}$ indicates successful execution without runtime errors, $\mathbb{I}_{valid}$ checks if the execution result $R_i$ is non-empty and meaningful (excluding empty lists or nulls), and $\mathbb{I}_{consist}$ assesses the completeness of the NL-MQL pair. We set the weights $\mathbf{w} = [0.3, 0.3, 0.2, 0.2]$. A sample is retained only if $\max_{y \in \mathcal{C}} S(y) \geq \tau$, where $\tau=0.8$ is the quality threshold. The complete procedure is summarized in Algorithm~\ref{alg:synthesis}.

\begin{algorithm}[ht]
\caption{Execution-Guided Data Synthesis Pipeline}
\label{alg:synthesis}
\begin{algorithmic}[1]
\Require Database $\mathcal{D}$, LLM $\mathcal{M}$, Target Count $T$, Threshold $\tau$
\Ensure Synthesized Dataset $\mathcal{D}_{syn}$
\State $\mathcal{D}_{syn} \leftarrow \emptyset$
\State $S_{schema} \leftarrow \text{ExtractSchema}(\mathcal{D})$
\State $S_{data} \leftarrow \text{SampleRepresentativeDocs}(\mathcal{D})$
\While{$|\mathcal{D}_{syn}| < T$}
    \State $c \leftarrow \text{SelectComplexity}(\{\text{Simple, Mod, Complex}\})$
    \State $d_{ctx} \leftarrow \text{ConstructPrompt}(S_{schema}, S_{data}, c)$
    \State $\mathcal{C} \leftarrow \emptyset$
    \For{$k \leftarrow 1$ \textbf{to} $N$} \Comment{Rejection Sampling Loop}
        \State $(r_k, \text{nl}_k, \text{mql}_k) \leftarrow \mathcal{M}.\text{Generate}(d_{ctx})$ \Comment{With CoT}
        \State $res_k \leftarrow \text{Execute}(\text{mql}_k, \mathcal{D})$
        \State $score_k \leftarrow \text{CalculateScore}(\text{mql}_k, res_k)$
        \State $\mathcal{C}.\text{add}(\{\text{nl}_k, \text{mql}_k, score_k\})$
    \EndFor
    \State $y^* \leftarrow \arg\max_{y \in \mathcal{C}} (y.score)$
    \If{$y^*.score \geq \tau$}
        \State $\mathcal{D}_{syn}.\text{add}(y^*)$
    \EndIf
\EndWhile
\State \Return $\text{CleanAndDeduplicate}(\mathcal{D}_{syn})$
\end{algorithmic}
\end{algorithm}

\end{document}